\documentclass[10pt,conference,compsocconf]{IEEEtran}
\usepackage{times}

\usepackage{caption}
\usepackage{url}
\usepackage{times}
\usepackage{xcolor}
\usepackage{tikz}
\usetikzlibrary{shapes}
\usepackage{graphicx}
\usepackage{orcidlink}
\usepackage{cleveref}
\usepackage{url}

\usepackage{breakurl}
\usepackage{hyperref}
\hypersetup{hidelinks,unicode,breaklinks}

\DeclareRobustCommand*{\IEEEauthorrefmark}[1]{%
	\raisebox{0pt}[0pt][0pt]{\textsuperscript{\footnotesize #1}}%
}

\usepackage{ifpdf}
\ifpdf
\pdfoutput=1 % we are running pdflatex
\pdftrue
\fi

\makeatletter
\def\ps@IEEEtitlepagestyle{
	\def\@oddfoot{\mycopyrightnotice}
	\def\@evenfoot{}
}
\def\mycopyrightnotice{
	{\footnotesize
		\begin{minipage}{0.8\textwidth}
			\centering
			Please cite as: Natascha Stumpp, Doris Aschenbrenner, Manuel Stahl, and Andreas Aßmuth, ``PLASMA~-- Platform for Service Management in Digital Remote Maintenance Applications,'' in \emph{Proc of the 10th International Conference on Cloud Computing, GRIDs, and Virtualization (Cloud~Computing~2019), Venice, Italy}, May 2019.
		\end{minipage}
	}
}
\makeatother
\makeatletter
\let\blx@rerun@biber\relax
\makeatother

\begin{document}

\pagenumbering{gobble}

\title{\textbf{\Large PLASMA~-- \underbar{Pla}tform for \underbar{S}ervice \underbar{Ma}nagement in\\[-1.5ex]Digital Remote
		Maintenance Applications}\\[0.2ex]}

\author{%
\IEEEauthorblockN{~\\[-0.4ex]\large Natascha Stumpp\IEEEauthorrefmark{1}, Doris Aschenbrenner\IEEEauthorrefmark{2}\,\orcidlink{0000-0002-3381-1673}, Manuel Stahl\IEEEauthorrefmark{3} and Andreas Aßmuth\IEEEauthorrefmark{4}\,\orcidlink{0009-0002-2081-2455}\\[0.3ex]\normalsize}
\IEEEauthorblockA{\IEEEauthorrefmark{1}ESSERT GmbH, Ubstadt-Weiher, Germany, Email: {\tt n.stumpp@essert.com}}
\IEEEauthorblockA{\IEEEauthorrefmark{2}Technische Universiteit Delft, Delft, Netherlands, Email: {\tt d.aschenbrenner@tudelft.nl}}
\IEEEauthorblockA{\IEEEauthorrefmark{3}Awesome Technologies Innovationslabor GmbH, Würzburg, Germany,\\ Email: {\tt manuel.stahl@awesome-technologies.de}}
\IEEEauthorblockA{\IEEEauthorrefmark{4}Technical University of Applied Sciences OTH Amberg-Weiden, Amberg, Germany, Email: {\tt a.assmuth@oth-aw.de}\\[1ex]}
}

\maketitle

\begin{abstract}
%\boldmath
To support maintenance and servicing of industrial machines, service processes are even today often performed manually and analogously, although supportive technologies such as augmented reality, virtual reality and digital platforms already exist. In many cases, neither technicians on-site nor remote experts have all the essential information and options for suitable actions available. Existing service products and platforms do not cover all the required functions in practice in order to map end-to-end processes. PLASMA is a concept for a Cloud-based remote maintenance platform designed to meet these demands. But for a real-life implementation of PLASMA, security measures are essential as we show in this paper.

\end{abstract}


\IEEEpeerreviewmaketitle

\section{Introduction}
% no \IEEEPARstart
A major competitive factor for manufacturing companies is a high and reliable availability of their production facilities. Despite already existing technology like Augmented Reality (AR) or Virtual Reality (VR), which has the potential to improve the service processes, a lot maintenance even today happens manually involving expert personnel. 
\begin{figure}[htbp]
	\centering%
	\small
	\begin{tikzpicture}
		\node (pict1) at (-1, 0) {\includegraphics[width=1cm]{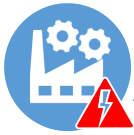}};
		\node (pict2) at (1, -1) {\includegraphics[width=1cm]{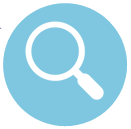}};
		\node (pict3) at (-1, -2) {\includegraphics[width=1cm]{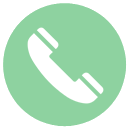}};
		\node (pict4) at (1, -3) {\includegraphics[width=1cm]{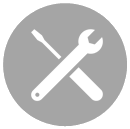}};
		\node (pict5) at (-1, -4) {\includegraphics[width=1cm]{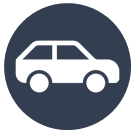}};
		\node (pict6) at (1, -5) {\includegraphics[width=1cm]{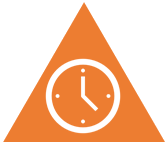}};
		\draw[thick, -latex] (pict1) -- (pict2);
		\draw[thick, -latex] (pict2) -- (pict3);
		\draw[thick, -latex] (pict3) -- (pict4);
		\draw[thick, -latex] (pict4) -- (pict5);
		\draw[thick, -latex] (pict5) -- (pict6);
		\node[anchor=west] at (-0.25, 0) {Malfunction of system};
		\node[anchor=east] at (0.25, -1) {On-premise troubleshooting};
		\node[anchor=west] at (-0.25, -2) {Contact support hotline};
		\node[anchor=east] at (0.25, -3) {Classic remote maintenance};
		\node[anchor=west, text width=3.25cm, text centered] at (-0.25, -4) {Expert has to travel to damaged system};
		\node[anchor=east] at (0.25, -5) {Loss of time and costs};
	\end{tikzpicture}
	\setlength{\belowcaptionskip}{-9pt}
	\caption{Course of actions without an intelligent maintenance platform.}\label{fig:maintenance}
\end{figure}
The common course of actions is depicted in Figure~\ref{fig:maintenance}. Imagine that production in a company suddenly succumbs because one of their machines stops working. At first, the workers try to find the reason for the malfunction themselves. Maybe, the company employs their own technicians for the maintenance of their systems. In this case, the workers call for one of these technicians. In most cases, these technicians do not have the same knowledge of the machine that specialists employed by the manufacturer of the machines have. In many cases, neither the technician nor the worker have all essential information or know about possible actions to solve the problem the right away. Therefore, if the technicians are not able to solve the problem, e.g., they cannot find a solution in the manual of the machine, the company contacts the manufacturer using their hotline or website. This is when classic remote maintenance comes into play. If the machine is connected to the Internet, one of the manufacturer's specialists connects to the system, e.g., via VPN, and tries to gather more information about the malfunction. There are numerous cases in which one of the specialists has to travel to a broken machine to repair it in on-site. An essential part of the machine might be physically broken and only the manufacturer is capable of installing a spare part. Assuming the manufacturer is situated in Europe and the company with the broken system is, e.g., in Australia, the travel might take days causing high costs for the company due to the outage.\par 
A small or medium-sized company today faces the challenge to implement their whole digital service processes in their existing environment, but only the currently available solutions usually cover just a small number of isolated use cases. Additionally, even though there is a large variety of such very specialised services, encapsulated platforms or IoT solutions readily available it is difficult to choose the ones the company really needs and that can be used in combination with services for other partial tasks of their digital service processes. For a complete mapping of application-driven end-to-end processes, it is necessary to realise a combination of these different platforms for small and middle-sized businesses which could probably struggle with the implementation by themselves. And these different platforms in practice do not necessarily interact properly with each other.

\subsection{Objective}
The joint project PLASMA aims for a holistic solution, which complements existing end-to-end business processes and supports the development of new service concepts, e.g., pay-per-x or x-as-a-service. Within the project an intelligent linkage between systems and platforms will be developed to allow integrated support and innovative business models all around service for production processes and facilities.\par 
The solution should seamlessly fit into all process models and should be integrable into existing system landscapes as well as Enterprise Resource Planning (ERP) systems. Additionally, PLASMA contains an information and knowledge management component to store and document instructions, tutorials, service reports, master data and offers a device- and location-independent visualization of it. PLASMA enables the user to handle complex machine data and real-time simulations presented in an intuitive way. With AR- and VR-support it will be possible to offer almost real guidance for maintenance and service cases. The service management platform can connect customers and suppliers and is intended to reshape the whole transparent life cycle of a product without exposing sensible data.

\subsection{Related work}
Currently, there is a vast change within automation industry which is attributed to be the ``fourth industrial revolution"; although this name is mainly used in a European context, there are similar movements in the USA and Asia. \cite{liao2017past} The goal of all these approaches is nearly the same: Whereas information and communication technology has advanced rapidly in recent years, the discovered trends and possibilities shall be transferred, so that the production industry can benefit from it. Although electronics and network infrastructure have of course been used for a long time in an industrial production setting, it is important to realise that plants and production machines are high investment goods which go together with slower innovation cycles. This means that while in the customer off-the-shelf segment, this year’s ``new'' hardware or software will be already considered ``old'' in half a year (and eventually even out of stack in a very short time span), the production eco-system has a relatively long usage period of hardware and software.\par 
But what is exactly changing due to ``Industry~4.0''? Next to individualised production, the core issues of Industry~4.0 can be formulated according to \cite{zuehlke2010smartfactory} as the integration of Internet and networking systems, smart objects and human machine interaction. This already emphasises the need for higher security requirements. Internet and Cloud applications \cite{mell2011nist} come with the need to integrate production systems in larger network infrastructures or even in the common Internet. The latter is strengthened by the trend to enable new kinds of human-machine interaction: Bring Your Own Device (BYOD) and remote access on industrial infrastructure with the help of mobile devices can without doubt offer new services or help to decrease costs. But they are also prone to attack scenarios.\par 
The general challenges of cybersecurity are already widely known. According to the 2017 Global State of Information Security Survey \cite{gsiss2017}, at least $80\,\%$ of companies in Europe have experienced at least one incident in 2016 and the number increased by $38\,\%$ compared to the preceding year. At the same time, approximately $69\,\%$ of European companies have either no or only basic understanding of their exposure to cyber risks and small and medium-sized companies tend to pay a higher price for this than larger companies. \cite{eucybersecurity2018}\par 
This topic increasingly receives the necessary political attention, for example, within the currently discussed European legislation regarding cybersecurity and vulnerability reporting. The above mentioned surveys mainly focus on ``common" office and server infrastructure, although the current transition of the production industry towards ``Industry~4.0" opens a large field of additional vulnerabilities. At the latest, since the Stuxnet \cite{langner2011stuxnet} malware, the possibility of damage on industrial infrastructure through the Internet has received worldwide attention. In order to understand where additional concern of security research should focus on in the upcoming years, we provide an overview over the current changes within the production industry and the resulting possible vulnerabilities.\par 
Due to the above explained transformation towards ``Industry~4.0" a multitude of devices become connected to the common Internet; IBM estimates that the number will increase to 40 billion by 2020. \cite{ibmdevices} To conclude from the above remarks, it cannot be expected that those devices have a sufficient amount of security protection. Rather, a lot of devices might consist of old, most probably unpatched equipment, but are wired to critical infrastructures. Practical proof of this problem can be, for example, obtained with tools, which automatically detect and index Internet-facing industrial systems. The Shodan computer search engine \cite{bodenheim2014evaluation} has been successfully tested to be able to index and identify Programmable Logic Controllers (PLCs). As those devices are standard components of industrial machines, several thousand devices can be found. As they are automatically tested on the running firmware and indexed accordingly, known vulnerabilities can be exploited easily.\par 
In a 2015 overview, Sadeghi et al. \cite{sadeghi2015security} lists a couple of cyberattacks on IIoT (Industrial Internet of Things) and emphasize the fundamental difference between CPPS (Cyber-Physical Production System) compared to classical enterprise IT systems. In the tradeoff between security and availibility, the CPPS requirements are fundamentally different. They mention numerous possible attacks on intellectual property, product piracy. After providing an overview to different security architectures for CPS (Cyber-Physical System), the article concludes with the following statement: ``However, existing security solutions are inappropriate since they do not scale to large networks of heterogeneous devices and cyber-physical systems with constrained resources and/or real-time requirements.''\par 
The book ``Cybersecurity for Industry 4.0" \cite{thames2017cybersecurity} provides the technological foundations of cybersecurity for the production domain. It addresses existing threats caused by (A)~humans, (B)~technical insufficiencies, and (C)~physical attacks of the actual IoT hardware. \cite{di2007hardware}\cite{shahri2012tree}\par 
Recently, NIST published a draft with considerations for managing Internet of Things cybersecurity and privacy rights. \cite{NISTIR8228} The main challenges are seen to protect device security, protect data security and protect individual's privacy. The publication focusses on ``Internet of Things" in the sense explained above and does not cover specific production topics.\par 
Are companies already aware of this topic? In the 2018 Global State of Information Security Survey (GSISS), $81\,\%$ of the companies judge IoT to be a critical part of at least some of their businesses. But only $39\,\%$ of survey respondents are confident that they have established ``sufficient digital trust~--security, privacy and data ethics-- into their adoption of IoT". Furthermore, the replies from organisations using robotics or automation show that $40\,\%$ fear a disruption of operations due to a cyberattack on those systems.

\section{The PLASMA Approach}\label{sec:approach}
\begin{figure}[htbp]
	\centering%
	\small
	\begin{tikzpicture}
		\node (swm) at (0, 0) {\includegraphics[width=2cm]{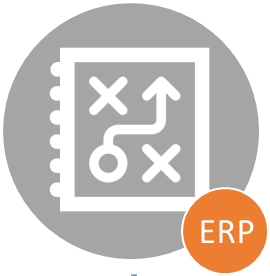}};
		\node (malf) at (-3.25, -0.25) {\includegraphics[width=1cm]{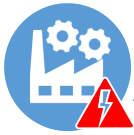}};
		\node (start) at (-1.85, -1.5) {\includegraphics[width=1cm]{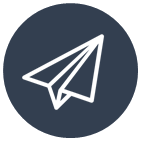}};
		\node (app) at (0, -2.25) {\includegraphics[width=1.8cm]{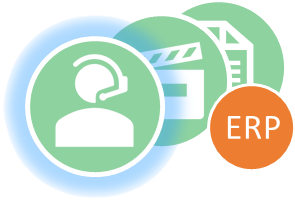}};
		\node (learn) at (1.85, -1.5) {\includegraphics[width=1cm]{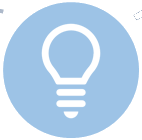}};
		\node (end) at (3.25, -0.25) {\includegraphics[width=1cm]{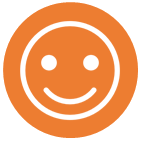}};
		\draw[thick, dashed] (swm) -- (malf) (swm) -- (start) (swm) -- (learn) (swm) -- (end);
		\draw[thick, -latex] (swm) -- (app);
		\draw[thick, -latex] (malf) -- (start);
		\draw[thick, -latex] (start) -- (app);
		\draw[thick, -latex] (app) -- (learn);
		\draw[thick, -latex] (learn) -- (end);
		\node at (0, 1.25) {Smart Workflow Manager};
		\node at (-3.25, 0.5) {Malfunction};
		\node[anchor=east, text width=2.5cm, text centered] at (-1.75, -2.4) {Troubleshooting is triggered};
		\node[text width=3cm, text centered] at (0, -3.25) {Problem solving with matching app};
		\node[anchor=west, text width=2.5cm, text centered] at (1.75, -2.4) {System learns from malfunction};
		\node at (3.25, 0.5) {Problem solved};
	\end{tikzpicture}
	\setlength{\belowcaptionskip}{-6pt}
	\caption{PLASMA workflow integrated in business processes}\label{fig:plasma_maint}
\end{figure}
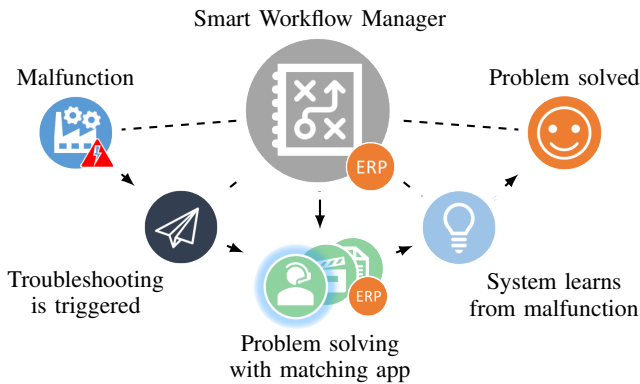

To implement a holistic interactive support for service processes in production environments with the goal to reduce time- and resource-consuming error search and troubleshooting it is necessary to evaluate the following features:
\begin{enumerate}
	\item Autonomous or automated event reporting in case of malfunction with digital communication tools like messengers or automated ticket systems,
	\item Automated delivery of context-sensitive data sheets, videos, reports, statistics or other helpful stored information on a large variety of devices with different presentation models (textual, 2D, 3D, virtual, augmented, simulated, etc.) 
	\item An interactive remote support assistance with a far-off specialist,
	\item A gateway to existing online-shop systems to automate the procurement of spare parts, and finally,
	\item A complete connection to well-known ERP and Customer Relationship Management (CRM) systems.
\end{enumerate}
With these features we aim to solve common use cases like a malfunctioning robot within an industrial plant. The goal is to find concrete solutions to elaborate a use case shown in Figure~2. The malfunction triggers the troubleshooting progress and tickets are created in an instant. A smart workflow manager can classify the incident and is able to suggests a solution depending on the severity of the error and archived data. The on-site worker gets useful information like data sheets, log files, instruction videos, virtual representations etc. to solve the issue by himself or receives remote support from a far-off specialist. All progress is documented and serves as new input for the smart workflow manager to sharpen its classification and support skills (cf. Figure~\ref{fig:plasma_maint}). 

\section{Security challenges}
The amount of information, as well as the aggregation of information makes a remote maintenance platform like PLASMA a high-value target for attackers. Because of the key knowledge on technologies, machines and algorithms stored in the system, economic espionage funded by competitors certainly is an issue. In case the attacker is not capable of extracting the desired information from the platform, for example, he could also try to bring the system down using a Distributed Denial of Service (DDoS) attack. This would lead to high financial losses for the providers of the platform and the customers relying on the system alike. Organised crime should also be taken into account because these attackers could also try to bring the system down and demand ransom money to be paid. Last but not least, secret services might become attackers, too, if the information stored in the platform is essential for companies or industrial branches in that country.\par 
To put it in a nutshell: since the remote maintenance platform is intended to be hosted in the Cloud, all of the already known security issues of Cloud services, e.g., documented by the Cloud Security Alliance in \cite{csa2017}, apply to PLASMA as well. The necessity to keep the platform available and accessible has already been stated. Considering additional security services, e.g., as recommended by CCITT X.800 \cite{x800}, it can be stated that their importance for the system security of PLASMA is equally essential:\par 
\textbf{Authentication:} It must be ensured that every entity communicating with the platform is properly authenticated. This means, the capability to perfectly identify users as well as attached machines is needed in order to prevent Spoofing or masquerading attacks.\par 
\textbf{Access Control:} In addition to authentication it must be ensured that authenticated users and machines alike are only able to access data they are allowed to. Due to the involvement of many different companies and roles, Role-based Access Control (RBAC) systems that have been adapted for use in Cloud environments, as proposed by Tang et al. \cite{cloud-rbac1} or Balamurugan et al. \cite{cloud-rbac2}, seem to meet this demand.\par 
\textbf{Confidentiality:} For big remote maintenance platforms, it seems likely that they will have competing companies as customers. This means, all data must be kept confidential such that, for instance, one company cannot get access to data from its competitor. As stated before, a remote maintenance platform stores and aggregates different types of information, like algorithms, procedures, etc., from manufacturers and customers or machine data about outages and errors. The system potentially gathers data that is relevant concerning the EU General Data Protection Regulation (GDPR), like working hours of operators or maybe errors made by certain operators. If technicians or experts use smartglasses during the error searching process, it is possible that other personnel might get recorded as well. This must be considered when it comes to GDPR-compliant saving of the data.\par
\textbf{Integrity:} PLASMA is intended to learn from previous errors and outages and if a malfunction occurs it is supposed to automatically suggest the most suitable action to deal with this scenario. An attacker might want to tamper with data in a way that leads to wrong suggestions, either to derogate trust in the remote maintenance platform or to harm an affected company. Other targets might be stored sensor data that lead to wrong simulation results when modified or falsified documentation on machines or manuals which could mislead technicians in case of a malfunction and cause even greater (physical) damage to the machine. Weir, A\ss muth and J\"ager have proposed strategies for intrusion monitoring in Cloud services and for managing forensic recovery in the Cloud. \cite{cloudjournal} It is planned to realise and evaluate these concepts for the remote maintenance platform.\par 
\textbf{Nonrepudiation:} It must be ensured that no party is capable of denying its involvement in any communication with or in the system. One reason to keep track of all actions in the system is to monitor the security of the system itself. But, of course, the provider of a remote maintenance platform wants to earn money with the system, too. Depending on the chosen business model the amount of messages or communication in general could be a metric to measure the usage of the system by a certain company and this may be used for billing.\par 
In order to emphasise the necessity for appropriate security measures in a Cloud-based remote maintenance platform, we revisit the use case described in Section~\ref{sec:approach} and depicted in Figure~\ref{fig:plasma_maint}. Obviously, the Cloud-based remote maintenance platform needs to be protected against DDoS attacks, otherwise the system would not take notice of the malfunctioning robot in one of the customer's industrial plants. The triggering of the troubleshooting process might be related to another security issue. Imagine the situation that there is no malfunctioning robot, but the troubleshooting is triggered by a manipulated sensor. The attacker might want to stop production in the industrial plant or learn how the maintenance platform deals with such problems. The adversary might also try to tamper with the smart workflow manager which could lead to inappropriate solutions for detected malfunctions and eventually cause even greater damage. In addition to that, if information about malfunctions and errors, manuals or machine data gets manipulated, the system will not be capable of learning properly how to handle such issues. Less knowing technicians working in the industrial plant but also specialists might be tricked into wrong actions. Security is essential for a system like PLASMA.

\section{Involved Partners}
The project core team consists of four parties: two industrial partners and two partners from academia.\par 
ESSERT GmbH provides its multi-user remote support system and large user base as an important starting point for the development. It already offers a detailed user and permission administration, generates service reports for further documentation and is available for iOS, Android devices and smartglasses. \cite{essertpage}\par 
Awesome Technologies is involved in a couple of Industry 4.0 projects which use Augmented and Virtual Reality with actual off-the shelf head-mounted displays, which also involves localization issues.\par
The cooperative setting of remote support is a very interesting topic within the framework of human supervisory control of smart cyber-physical production systems (smart factory) at TU Delft.\par 
The research group of Prof. Dr. A\ss muth at OTH Amberg-Weiden has been working on concepts and solutions to ward off cyber-attacks aimed specifically at production facilities or vehicles for many years. In cooperation with international colleagues, concepts for increasing the security of Cloud services and securing forensic data in the Cloud have been published as well. \cite{cloudjournal}\par 
The mentioned partners are currently looking for additional partners and funding programs for a PLASMA funding proposal.

\section{Conclusion and Future Work}
To compete on Cloud service markets SMEs need to focus on security challenges. Launching a great idea on the market may fail due to insufficient data security or privacy issues. Meeting a customer’s high expectations for security is essential and a great challenge for SMEs because there are no negotiation opportunities. The authors are convinced that a Cloud-based remote maintenance platform, like PLASMA, will be needed in future. Therefore, they plan to realise such a system in a funded research project as a collaboration of industrial partners and partners from academia.

\end{document}